\documentclass[]{spie}  

\usepackage{amsmath,amsfonts,amssymb}
\usepackage{graphicx}
\usepackage[colorlinks=true, allcolors=blue]{hyperref}
\title{Optimizing Fourier-Filtering WFS to reach sensitivity close to the fundamental limit}

\author[a]{V.Chambouleyron}
\author[c]{O. Fauvarque}
\author[d]{C. Plantet}
\author[b,a]{J-F. Sauvage}
\author[a,b]{N. Levraud}
\author[a,b]{M. Cissé}
\author[a]{B. Neichel}
\author[b,a]{T. Fusco}

\affil[a]{Aix Marseille Univ, CNRS, CNES, LAM, Marseille, France}
\affil[b]{DOTA, ONERA, Université Paris Saclay, F-91123 Palaiseau, France}
\affil[c]{IFREMER, Laboratoire Detection, Capteurs et Mesures (LDCM), Centre Bretagne, ZI de la Pointe du Diable, CS 10070, 29280, Plouzane, France}
\affil[d]{INAF - Osservatorio Astronomico di Arcetri}

\authorinfo{Futher author information: vincent.chambouleyron@lam.fr}

\begin{document} 
\maketitle

\begin{abstract}
To reach the full potential of the new generation of ground based telescopes, an extremely fine adjustment of the phase is required. Wavefront control and correction before detection has therefore become one of the cornerstones of instruments to achieve targeted performance, especially for high-contrast imaging. A crucial feature of accurate wavefront control leans on the wavefront sensor (WFS). We present a strategy to design new Fourier-Filtering WFS that encode the phase close from the fundamental photon efficiency limit. This strategy seems promising as it generates highly sensitive sensors suited for different pupil shape configurations.  
\end{abstract}


\section{INTRODUCTION}

\label{sec:intro}
Adaptive optics (AO) is a technique allowing to compensate for the impact of atmospheric turbulence on telescopes that has become essential for a large number of astrophysical applications. Motivated in particular by the hunt for exoplanets \cite{2012SPIE.8447E..0BK}, AO systems pushing the limits of performance are currently being developed, called extreme AO system (XAO). These systems relies on high-order deformable mirrors with fast real time computation. The fundamental limit of such instruments is based on the quality of the measurements provided by the optical device at the heart of the AO system: the wavefront sensor (WFS). One key aspect driving the XAO instruments is the WFS sensitivity, that can limit the number of controlled modes and the speed of the loop. The Fourier Filtering WFS (FFWFS) represents a wide class of sensors of particular interest thanks to their superior sensitivity. From a general point of view, a FFWFS consists of a phase mask located in an intermediate focal plane which performs an optical Fourier filtering. This filtering operation allows the conversion of phase information at the entrance pupil into amplitude at the pupil plane, where a quadratic sensor is used to record the signal \cite{2004OptCo.233...27V, Guyon_2005, fauvOptica}. The goal of this paper is to optimise, in the sense of the highest possible sensitivity, the shape of the critical optical component of a FFWFS: its focal plane filter device. To that end, we first present in section 2 a noise propagation model for FFWFS allowing to define quantitatively what we call their sensitivities. In section 3, we present a new way to generate highly sensitive masks thanks to a numerical optimization of their shape.

\section{A noise propagation model for all FFWFS}

A new noise propagation model was developed for all kind of FFWFS \cite{Chambou2021}. This model, that was derived in the small phase approximation, allows to link the estimation error $\sigma$ on a given mode due to noise propagation with the number of photons $N_{ph}$ available for the measurement. This relationship is described through the following formula:

\begin{equation}
\sigma^{2}_{\phi_{i}} = \frac{N_{sap}\times\sigma_{ron}^{2}}{ s^{2}(\phi_{i})\times N_{ph}^{2}}+\frac{1}{s^{2}_{\gamma}(\phi_{i})\times N_{ph}}
\label{eq:ch4:propagation_bruit_mesure}
\end{equation}

Where $\phi_{i}$ is the considered mode, $N_{sap}$ is the number of sampling points in the pupil, $\sigma_{RON}^{2}$ the detector RON, and finally $s$ and $s_{\gamma}$ are the so-called sensitivities. One can spot two part in this formula: the left-hand side is describing the RON noise propagation (it actually apply to any kind of uniform noise on the detector), and the right hand side corresponds to the photon noise propagation. For a given mode, the two sensitivity terms can be found using:

\begin{itemize}
\item The corresponding signal of the mode in the interaction matrix: $\delta I(\phi_{i})$.
\item The reference intensities $I_{0}$, i.e the intensities corresponding to the reference wavefront (we assume a flat wavefront for this study).
\end{itemize}

\begin{equation}
\label{sensibilité_mode_i_photon}
s(\phi_{i}) = \sqrt{N_{sap}}\times\big|\big|\delta I(\phi_{i})\big|\big|_{2}
\quad \text{and} \quad
s_{\gamma}(\phi_{i}) = \big|\big|\frac{\delta I(\phi_{i})}{\sqrt{I_{0}}}\big|\big|_{2}
\end{equation}

These two sensitivities have no reason to be equal. Therefore, when comparing two FFWFS, it is required to compare both sensitivities in order to have a fair comparison. In the rest of this paper, we will call $s$ simply “sensitivity” because it actually corresponds to the sensitivity for any uniform noise on the detector (but the main one for our application is the RON) and we will call $s_{\gamma}$ the “sensitivity to photon noise”. These two sensitivities are bounded \cite{paterson,PhysRevApplied.15.024047}, and they can’t reach value above 2. This value of 2 set an upper fundamental limit for both sensitivities.

\begin{equation}
\begin{split}
    0\leq\ &s \leq 2\\
    0\leq\ &s_{\gamma} \leq 2
\end{split}
\end{equation}

The Zernike WFS (ZWFS) is often considered as the most sensitive WFS \cite{Guyon_2005}.  In a previous study, we used the sensitivity metrics introduced previously to assess performance of ZWFS class \cite{2021A&A...650L...8C}. It was shown that the classical ZWFS (dot diameter $p=1.06\ \lambda/D$ and phase depth $\delta = \pi/2$) is actually not the most sensitive sensor ($s = 0.9$ and $s_{\gamma} = 1.2$), and that its sensitivity can be improved by increasing the phase-dot diameter (we introduced the Z2WFS: $p=2\ \lambda/D$ and $\delta = \pi/2$). The aim of this paper is to find even more sensitive configurations for which $s$ and $s_{\gamma}$ are reaching values even closer from the maximum of 2. We will focus only on the sensitivity in the framework of the linear regime. Hence, we will not consider the dynamic behaviour of the generated FFWFS in this study.

\section{A new technique to create highly sensitive masks}

\subsection{Principle of the technique: convolutional model and numerical optimization}

The convolutional model is a model which describes any FFWFS measurement as a convolutional operation described by its impulse response and its associated transfer function. This model, introduced in \cite{fauvConv}, gives an interesting relationship between the shape of the filtering mask and the RON sensitivity map:

\begin{equation}
s \approx \sqrt{|\textbf{TF}|^2\star PSF}
\label{eq:sens_2}
\end{equation}

where $PSF$ corresponds to the Point-Spread Function in the plane of the filtering mask, and $\textbf{TF}$ is the transfer function of the considered FFWFS. This transfer function can be written as:

\begin{equation}
    \textbf{TF} = 2\text{Im}(m\star\overline{m\times PSF})
    \label{eq:ch2:TF_simple}
\end{equation}

where $m$ is the WFS filtering mask. Hence, we have a 2D sensitivity function $s$ which takes as argument two maps:\\
\begin{itemize}
    \item[$\bullet$] $PSF$ which corresponds to the Point-Spread Function at the filtering mask plane.
    \item[$\bullet$] $m$: the WFS filtering mask.\\
\end{itemize}

\begin{figure}[ht]
	\begin{center}
	\includegraphics[scale=0.47]{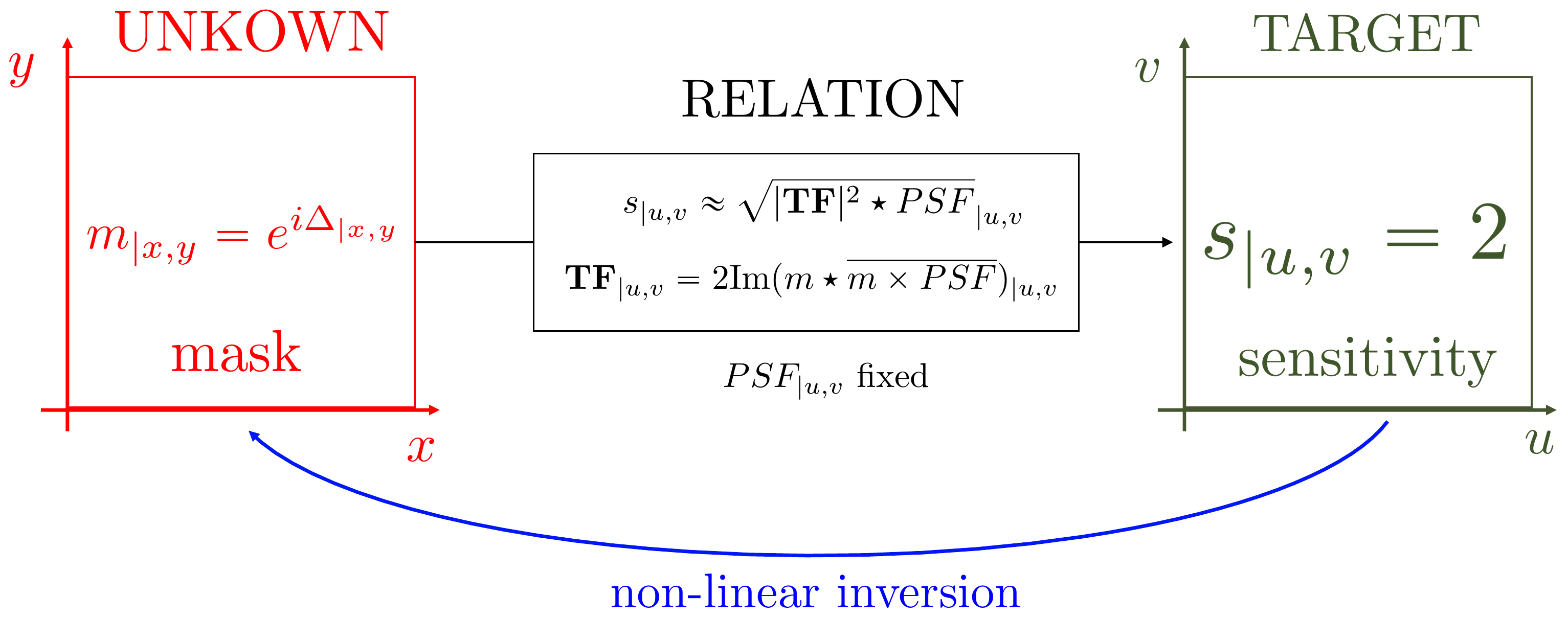}
	\caption{\textbf{Mask optimization.} The target sensitivity map is chosen to be maximal on a given spatial frequency range $(u,v)$.}
	\label{fig:ch4:optiScheme}
	\end{center}
\end{figure}

For a given pupil shape (and therefore a given PSF), we suggest here to numerically optimise the 2D sensitivity map through Eq. \ref{eq:sens_2} and Eq. \ref{eq:ch2:TF_simple}. This sensitivity map therefore depends only on the mask $m$: $s = s(m)$. We call $s_{target}$ the targeted sensitivity. We aim at inverting the problem thanks to numerical optimizations : given a targeted sensitivity, it is possible to compute the corresponding mask. This technique is illustrated in a schematic way figure \ref{fig:ch4:optiScheme}.

The least-square criterion is used for the optimization, and the scoring function $F$ (function to be minimized) is written:

\begin{equation}
\mathcal{F}(m) = ||s_{target}-s(m)||_{2}
\label{eq:ch4:score}
\end{equation}

$\mathcal{F}$ is a function which goes from $\mathbb{R}^{k}$ in 
$\mathbb{R}$ where $k$ is the number of parameters on which the mask is optimized. It is therefore possible to implement non-linear numerical optimization methods in order to find a minimum of this score function. However, it is not obvious that this function is convex: it can thus exhibit several local minima, and it will consequently be difficult to be ensured to have reached the global minimum at the end of an optimization process. To carry out this numerical optimization, we use the \textit{lsqnonlin} function of MATLAB which is a non-linear least squares problem solver using the \textit{Levenberg-Marquardt} algorithm. This algorithm is based on classical gradient descent methods. The goal of this study is to build a mask with an optimal sensitivity, so the target sensitivity map is chosen to reach a value of $2$ at each point in the frequency space: $s_{target} = 2$. Our score function thus becomes:

\begin{equation}
\mathcal{F}(m) = ||2-s(m)||_{2}
\label{eq:ch4:score_max}
\end{equation}

An important point that must be emphasized: the optimization will be done here on the sensitivity with respect to a uniform noise (like RON), because it is the one for which the convolutional model gives us an explicit formula. We will show later that this study could be easily applied to photon noise sensitivity.

\subsubsection*{Optimization area}

The way FFWFS are encoding the phase leads to a trade-off in the optimization: a gain in sensitivity for high frequencies comes with a loss of sensitivity for very low frequencies \cite{2021A&A...650L...8C}. Thus, the frequency range of the area over which we are optimizing has an importance: as we work with the least squares criterion, a target sensitivity extended over a large frequency range will tend to reduce the importance of low frequencies during optimization, and vice versa. One could also imagine working with always the same frequency range, and using weighted least squares as an optimization criterion.

\subsubsection*{Reduced parameters for mask}

The description of the mask $m$ is made with a number of points which depends on the resolution chosen in the focal plane and on its extension in this plane. Performing an optimization on all the phase points of a mask can be tedious in terms of computation time. Hence, strategies can be developed to reduce the parameters describing the mask: number of faces, angle, dot diameter for ZWFS, etc... We will use in particular the fact that we can consider the problem with symmetry of revolution and make a radial optimization , which greatly reduces the number of pixels in input and output. Of course, this strategy only applies for rotationally symmetric pupils and will produce optimized masks which are themselves rotationally symmetric. In general: for a given system , it is crucial to find symmetries allowing a reduction of the parameters to be optimized.

\subsection{Implementation of the technique through some examples}
\label{section_opti}

We start by setting up the optimization strategy in the context of a rotationally symmetrical pupil, of which we will study two variants: the full circular pupil and the obstructed circular pupil. As mentioned before, we take as target sensitivity, the maximum sensitivity $s_{target} = 2$. Moreover, only the phase masks $m=e^{i\Delta}$ (where $\Delta$ is a real function) are considered in order not to generate amplitude masks which would lead to a loss of photons. Finally, the score function is written, by squaring the sensitivity maps (Eq. \ref{eq:sens_2}, \ref{eq:ch2:TF_simple} and \ref{eq:ch4:score_max} ):

\begin{equation}
\mathcal{F}(\Delta) = ||4-|2\text{Im}\big(e^{i\Delta}\star (e^{-i\Delta}PSF)\big)|^2\star PSF||_{2}
\label{eq:ch4:score_max}
\end{equation}

For all optimizations, we generate a null phase mask $\Delta=0$ as a starting point for MATLAB's nonlinear optimization. We start the optimizations for the case of a PSF being a perfect Airy disk,corresponding to a full pupil.

\subsubsection{Full pupil case}

For this first optimization, we consider the case of a full pupil. The simulation parameters are as follows: $80$ pixels in the pupil with a sampling in the focal plane at twice the Shannon criterion (\textit{i.e.} $4$ pixels per $\lambda/D$).

\subsubsection*{Optimized mask characteristics}

\begin{figure}[ht]
	\begin{center}
	\includegraphics[scale=0.62]{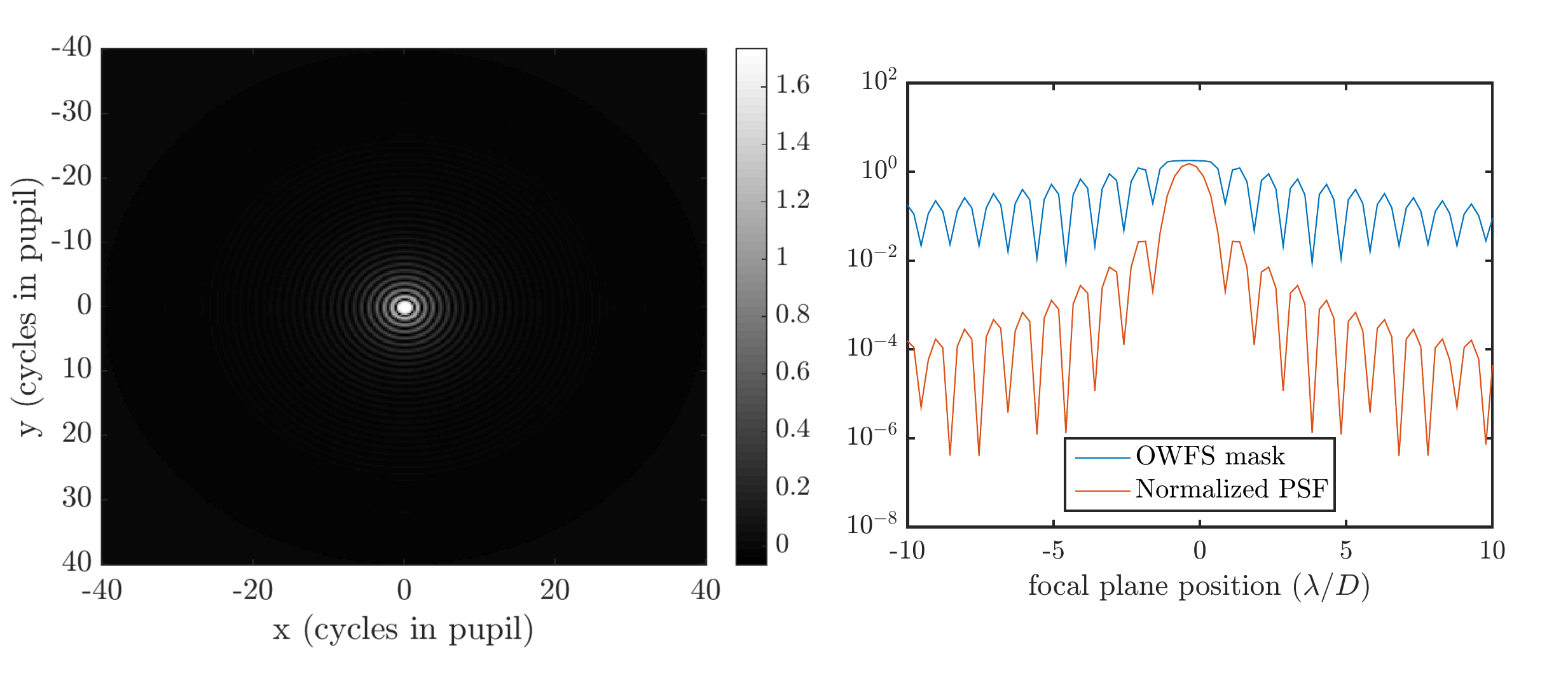}
	\caption{\textbf{Optimized mask for a full pupil configuration. Left:} $arg(m)$. \textbf{Right: } cut of the mask phase w.r.t normalized PSF- log scale. }
	\label{fig:ch4:OZWFS}
	\end{center}
\end{figure}

The optimization is done on all available spatial frequencies: i.e with a radius of $40\ \lambda/D$ in the focal plane. The obtained phase mask $\Delta$ is given figure \ref{fig:ch4:OZWFS}. The mask thus created is called the Optimized WFS (OWFS). It can be seen that this mask has a structure very similar to the PSF. Note here the similarity with ZWFS class: the mask has a phase-shifting zone in the center of a size of about $2\ \lambda/D$. But instead of being a sharp step, the central dot is somehow apodized. In addition, the created mask not only has a localized phase-shift dot in the center, but has also phase-shift zones that coincide with the energy distribution in the focal plane. This agreement between the phase shift zones of the mask and the energy distribution in the PSF is very visible on the cut provided in figure \ref{fig:ch4:OZWFS}-Right.

\subsubsection*{Reference intensities and response to phase aberrations}

\begin{figure}[ht]
	\begin{center}
	\includegraphics[scale=0.6]{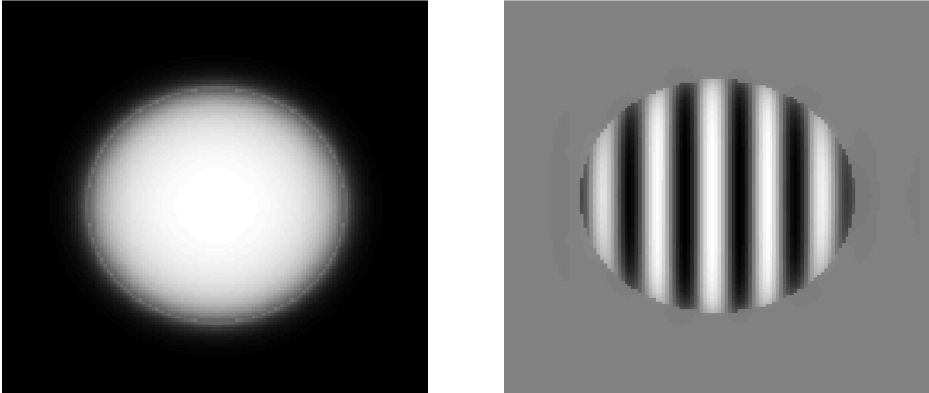}
	\caption{\textbf{Left:} reference intensities. \textbf{Right:} response to a cosine phase.}
	\label{fig:ch4:OWFS_reponse}
	\end{center}
\end{figure}

It is possible to look at the form of the response given by this mask. The reference intensities and the response to a cosine in the pupil are given in figure \ref{fig:ch4:OWFS_reponse}.

It can be seen that the response is very close to the one produced by the ZWFS class: the mask created is a "direct phase" sensor.

\begin{figure}[ht]
	\begin{center}
	\includegraphics[scale=0.7]{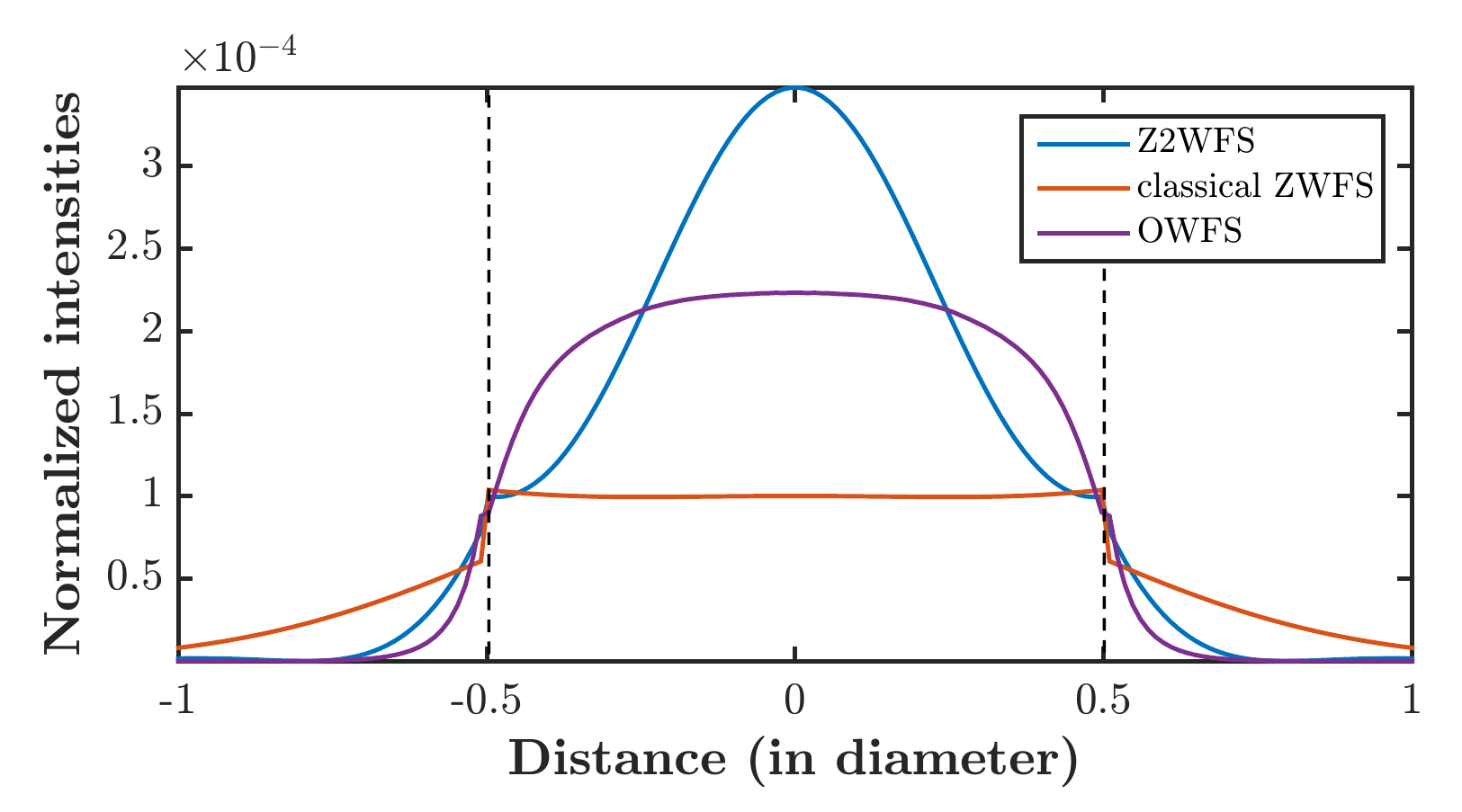}
	\caption{\textbf{Reference intensities cut for some FFWFS.} Pupil edges are the black dashed lines.}
	\label{fig:ch4:OWFS_cut}
	\end{center}
\end{figure}

\subsubsection*{Sensitivities}

\begin{figure}[ht]
	\begin{center}
	\includegraphics[scale=0.6]{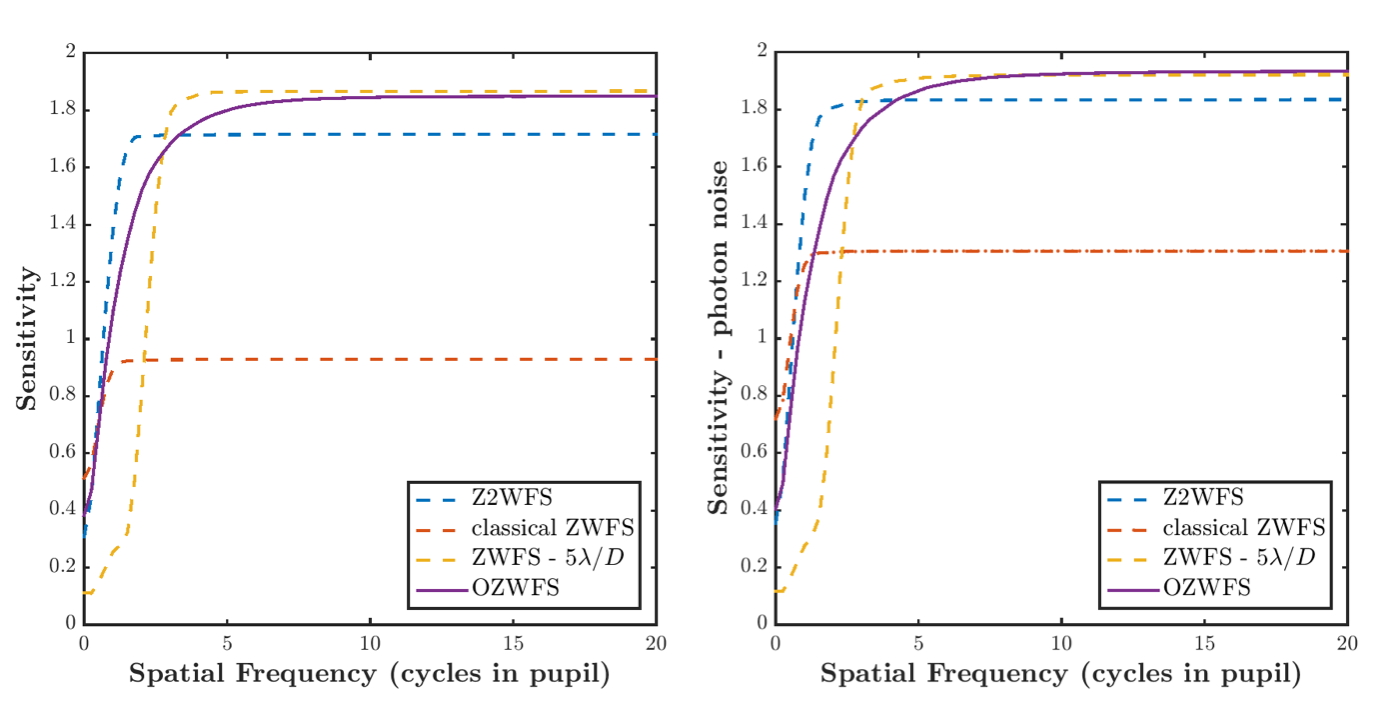}
	\caption{\textbf{Sensitivities curves for OWFS} compared to ZWFS class.}
	\label{fig:ch4:OZWFS_sensi}
	\end{center}
\end{figure}

We compare with end-to-end simulations the performance in sensitivities of the optimized mask generated with the ZWFS class. The sensitivities are plotted according to pure spatial frequencies. The results are given in figure \ref{fig:ch4:OZWFS_sensi}. We can see that the sensitivity of this WFS manages to reach the one of a ZWFS with a large dot (diameter of $5\ \lambda/D$ in yellow dotted lines on the figure) while still having great sensitivity at low frequencies (very similar to that of the Z2WFS, in blue dotted line). On the chosen frequency range, this sensor is therefore, to our knowledge, the most sensitive sensor ever built. The same behavior is also observed for the sensitivity to photon noise: like ZWFS, there is a correlation between the two types of sensitivities.

\begin{figure}[!ht]
	\begin{center}
	\includegraphics[scale=0.6]{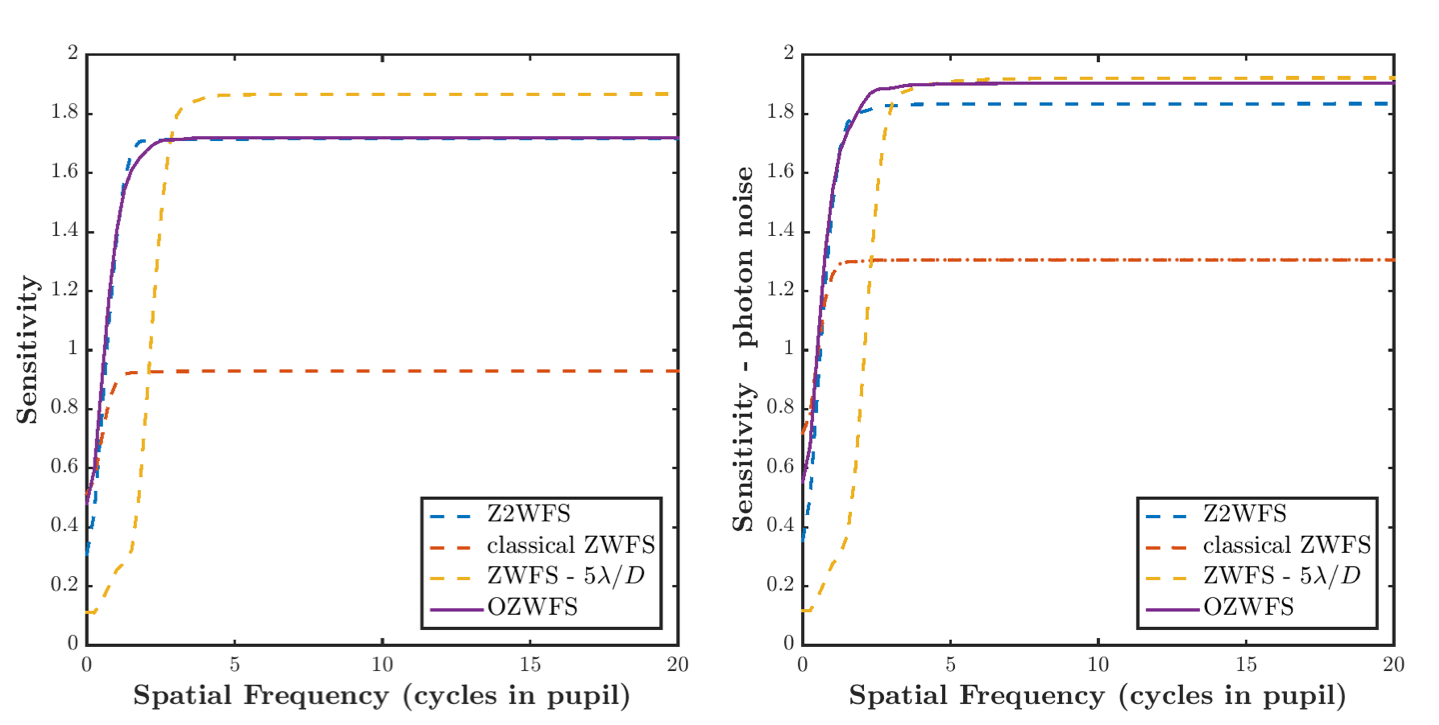}
	\caption{\textbf{Sensitivity curves for an optimization done on a smaller area in the space frequency} compared to ZWFS class.}
	\label{fig:ch4:OZWFS_sensi_LF}
	\end{center}
\end{figure}

We remind the reader that the result of the optimization in the sense of the least squares is sensitive to the zone in the space of the frequencies over which the optimization is done. It is thus possible to reduce this optimization zone or use a different weight on each frequencies, in order to make more use of the low frequencies in the optimization process. If we perform exactly the same optimization, but on the frequencies included in a radius of $5\ \lambda/D$ instead of $40\ \lambda/D$, we obtain the sensitivity curves figure \ref{fig:ch4:OZWFS_sensi_LF}. As expected, it can be seen that for the sensitivity to uniform noise (on which the optimization is carried out) the low frequencies have been enhanced with respect to the high frequencies. This time in terms of sensitivity for uniform noise, the sensor thus optimized equals the classic ZWFS for low frequencies while having a sensitivity equivalent to the Z2WFS for high frequencies: we have clearly overcome any counterpart on the low spatial frequencies. For photon noise, the sensitivity curve does not follow exactly the same trend as the one for uniform noise: this is not surprising, we know that this kind of sensitivity also depends on the structure of the reference intensities.

The optimization technique developed here is therefore very effective since it has made possible to develop a mask that is very sensitive to high frequencies while maintaining good sensitivity to low frequencies. In addition, it can be seen that the reference intensities produced by this mask are much more uniform than those produced in the case of the Z2WFS. Just like ZWFS class, the major drawback of this mask ultimately remains its very strong chromaticity and very small dynamic range.

\subsubsection{Obstructed pupil case}

For this study, the previous parameters are left unchanged, while the considered pupil has now a central obstruction of $30\%$ in diameter.

\begin{figure}[ht]
	\begin{center}
	\includegraphics[scale=0.3]{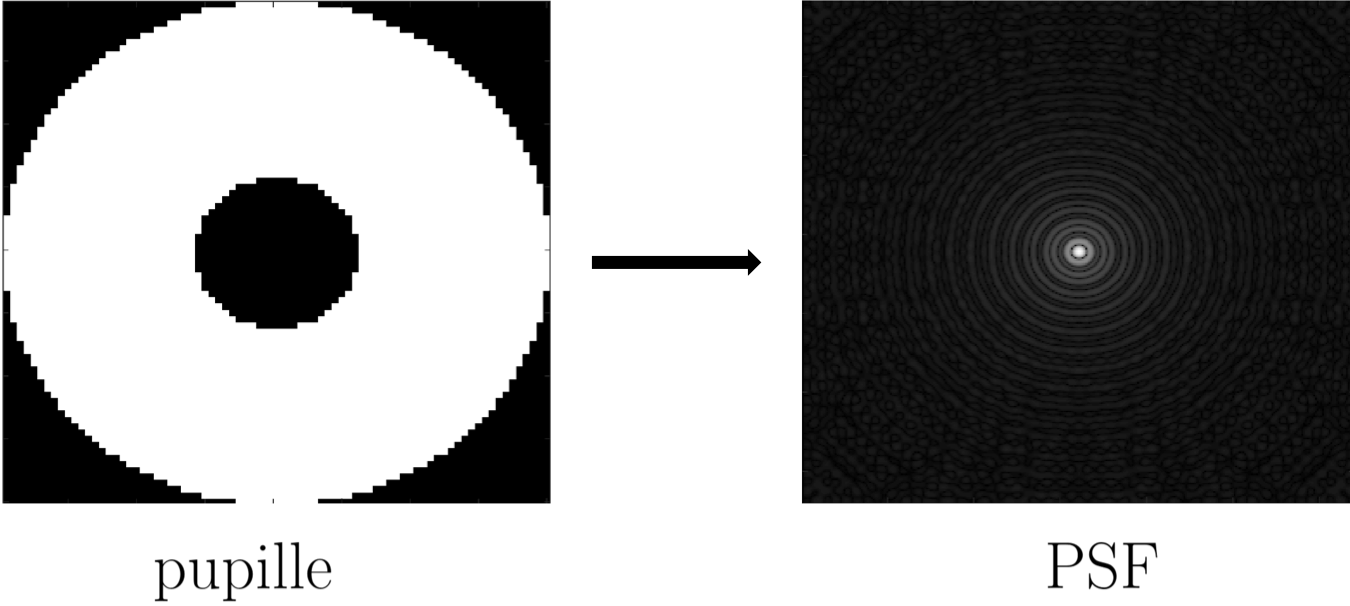}
	\caption{\textbf{Obstructed pupil of $30\%$ and corresponding PSF.}}
	\label{fig:ch4:pupille_trou}
	\end{center}
\end{figure}

\subsubsection*{Optimized mask characteristics}

\begin{figure}[ht]
	\begin{center}
	\includegraphics[scale=0.65]{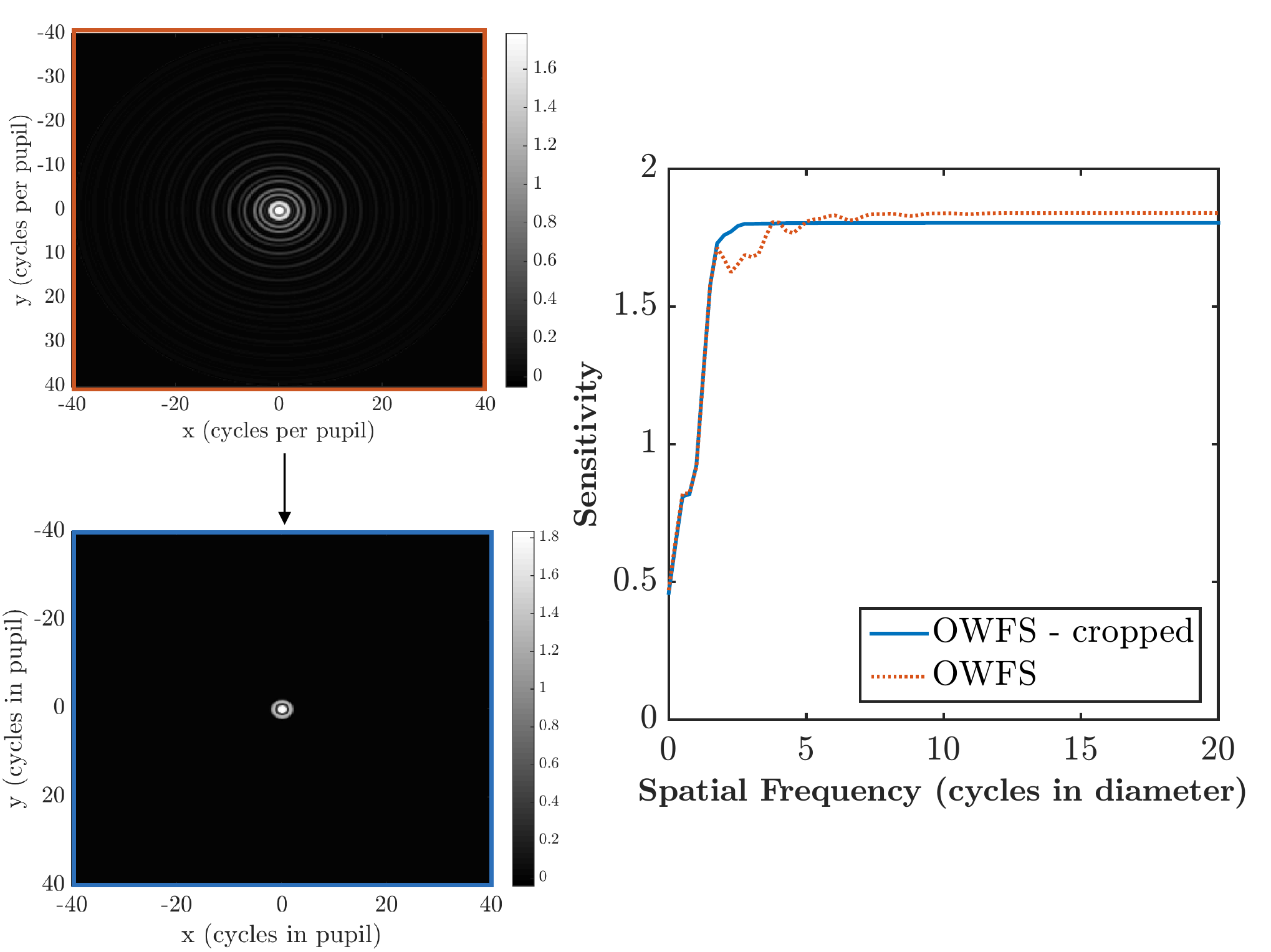}
	\caption{\textbf{Optimized masks for an obstructed pupil case.}}
	\label{fig:ch4:OWFS_masque_cropped}
	\end{center}
\end{figure}

After optimization, the obtained mask is presented in figure \ref{fig:ch4:OWFS_masque_cropped}. Once again, it is clearly observed that the optimization converges towards a mask where the phase-shift zones coincide with the energy distribution in the focal plane. In the case of an obstructed pupil, it is well known that the PSF exhibits a significant first ring: this structure is clearly found in the optimized mask.

With this mask, we see that the frequency sensitivity map shows rebounds around $5\ \lambda/D$ (figure \ref{fig:ch4:OWFS_masque_cropped}). This structure is produced by the counterparts far from the center of the mask formed during the optimization. We then decide to set the phase points of the mask far from the center to 0, in order to keep only its central part out of phase: it allow to cancel this erratic behavior around the frequencies $5\ \lambda/ D$, while leading to a small loss of sensitivity for high frequencies. This operation is quite similar to optimizing over a smaller frequency range. The resulting mask has a smoother shape.

\subsubsection*{Reference intensities and response to phase aberrations}

\begin{figure}[ht]
	\begin{center}
	\includegraphics[scale=0.6]{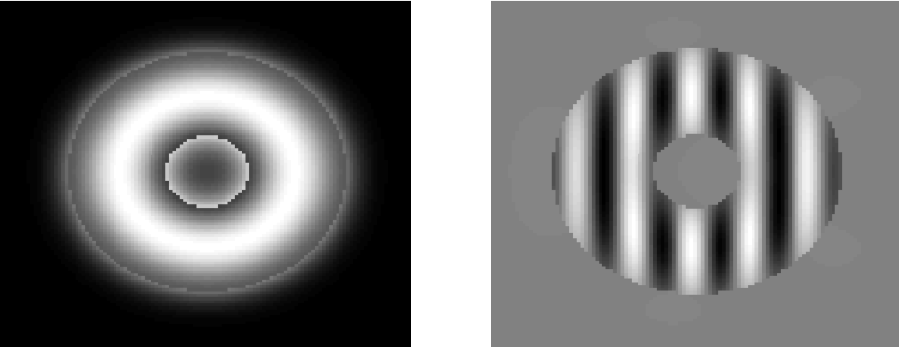}
	\caption{\textbf{Left:} Reference intensities. \textbf{Right:} response to a cosine phase.}
	\label{fig:ch4:OWFS_cropped_reponse}
	\end{center}
\end{figure}

The reference intensities and the cosine response for this mask are given in figure \ref{fig:ch4:OWFS_cropped_reponse}.

\begin{figure}[ht]
	\begin{center}
	\includegraphics[scale=0.65]{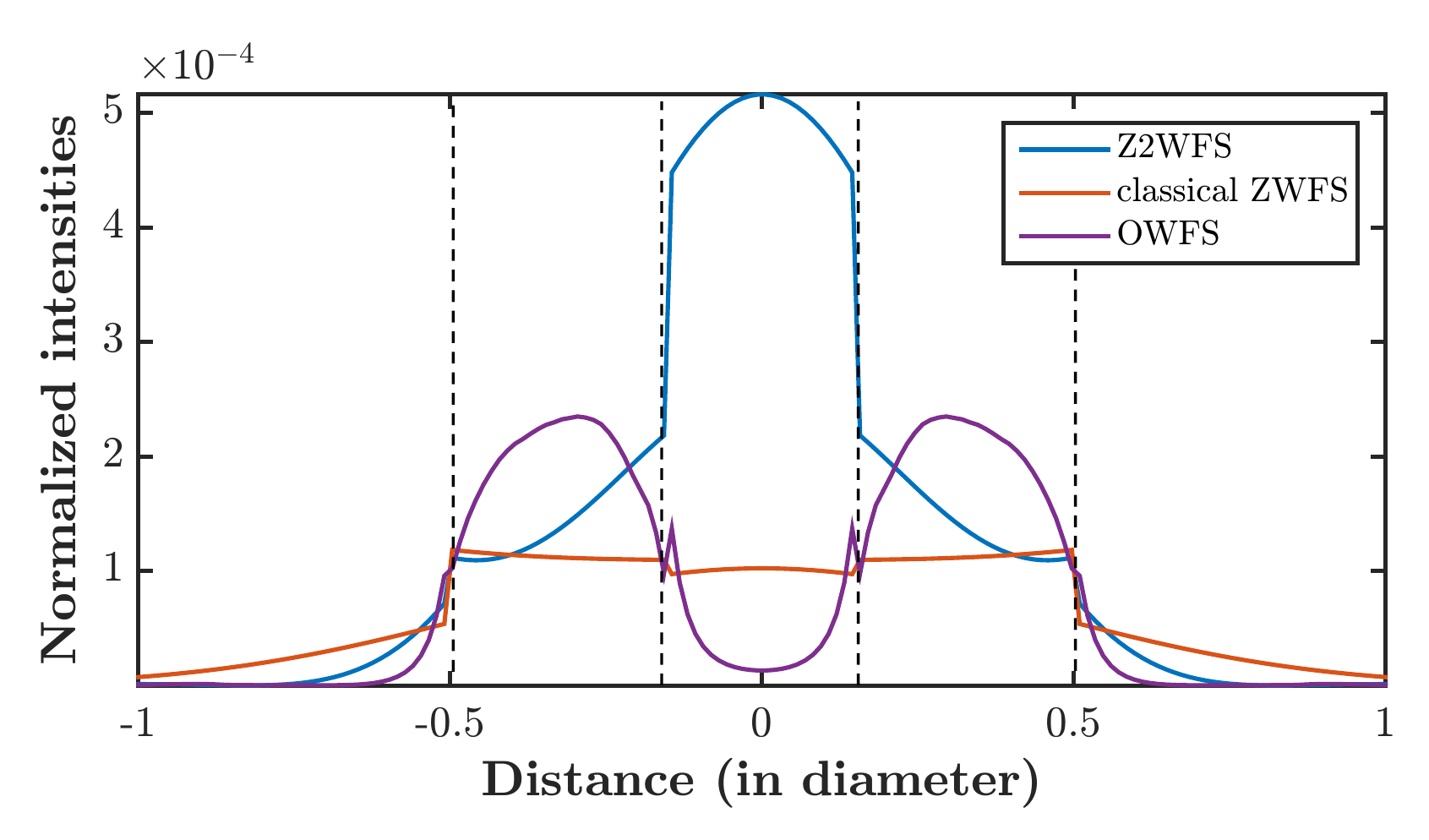}
	\caption{\textbf{Reference intensities cut for some FFWFS for an  obstructed pupil case.} Pupil edges are the black dashed lines.}
	\label{fig:ch4:OZWFS_cut_cropped}
	\end{center}
\end{figure}

The cut of the reference intensities (figure \ref{fig:ch4:OZWFS_cut_cropped}) of the mask reveals a very interesting behavior: contrary to the case of the Z2WFS or the classic ZWFS, only a small part of photons are present in the central obstruction. However, in the case of ZWFS class, it is recalled that the support of the linear behaviour is the pupil footprint. Thus all the photons located in the central obstruction do not take part in the analysis. Therefore OWFS for an obstructed pupil already seems to have some advantages just by looking at its reference intensities.

\subsubsection*{Sensitivities}

\begin{figure}[ht]
	\begin{center}
	\includegraphics[scale=0.6]{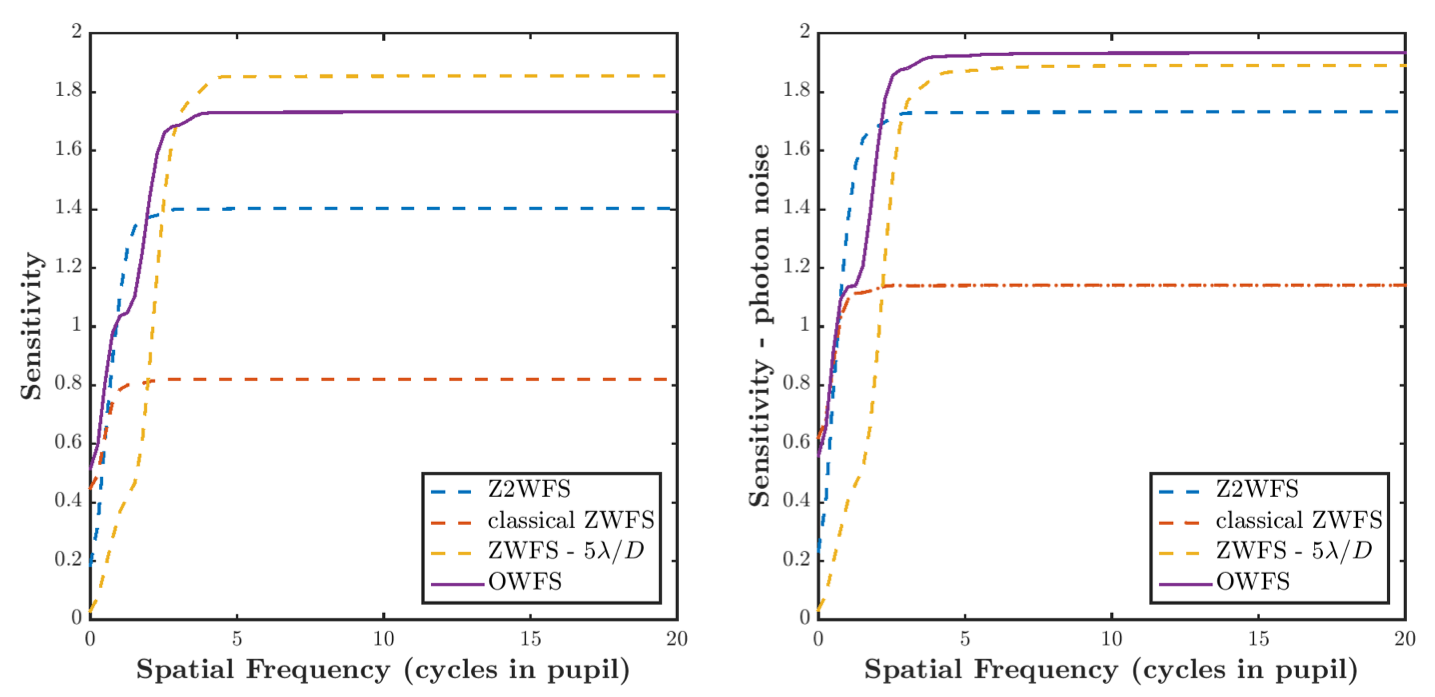}
	\caption{\textbf{Sensitivity curves for the optimized mask in the case of an obstructed pupil} compared to ZWFS class.}
	\label{fig:ch4:OWFS_cropped_sensi}
	\end{center}
\end{figure}

As presented before, let's compare our new sensor to ZWFS class (figure \ref{fig:ch4:OWFS_cropped_sensi}). The gain observed in the previous full pupil case is also found here. The strong gain for the sensitivity in the case of the optimized mask is explained by the extra photons available in the pupil footprint compared to the case of ZWFS class, for which there is a loss of energy in the central obstruction.

All large telescopes having quite large central obstructions, the case studied here is quite important: it defines \textbf{a new form of reference mask for high sensitivity wavefront sensing on this kind of pupil}.

\subsubsection{More complicated pupils}

\begin{figure}[ht]
	\begin{center}
	\includegraphics[scale=0.6]{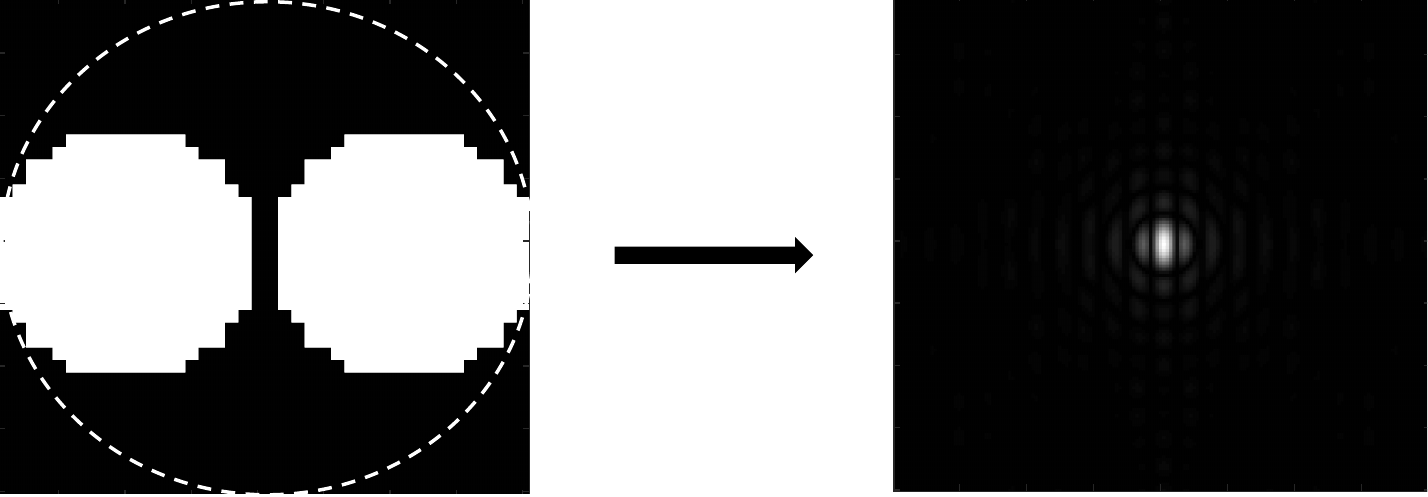}
	\caption{\textbf{LBT-style pupil and corresponding PSF.}}
	\label{fig:ch4:pupil_LBT}
	\end{center}
\end{figure}

As demonstrated before, our optimization algorithm can be adapted to different pupil shapes. In the previous cases, the symmetry of revolution was of great help in terms of computation time. Nevertheless, we can now optimize our masks for pupils with more complicated shapes, where the symmetry of revolution is no longer available. We choose an example: a pupil in the like the \textit{Large-Binocular Telescope} (LBT), i.e. composed of two juxtaposed primary mirrors (figure \ref{fig:ch4:pupil_LBT}) \cite{2012SPIE.8444E..1AH}.

\begin{figure}[ht]
	\begin{center}
	\includegraphics[scale=0.8]{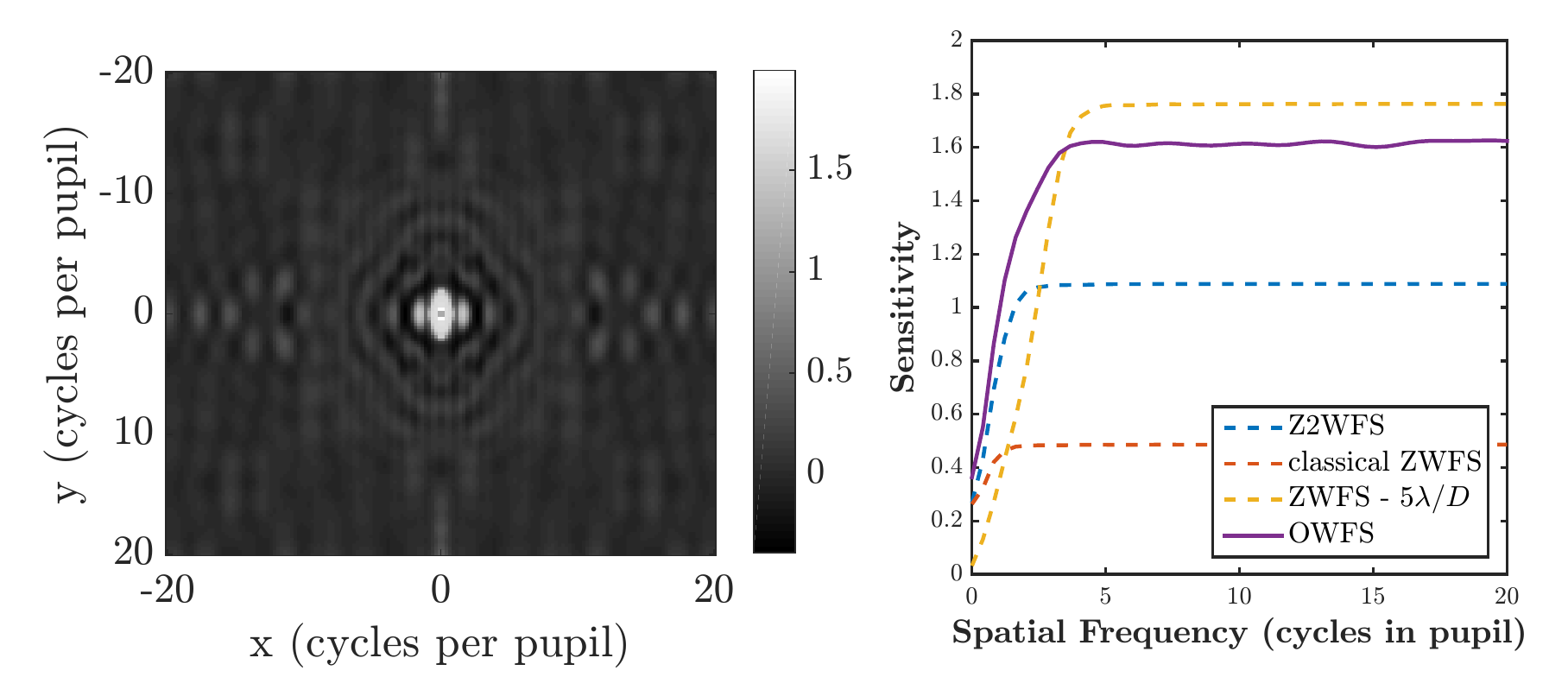}
	\caption{\textbf{OWFS mask shape for LBT-like pupil and corresponding sensitivity curve}. Once again, the PSF structure can be seen in the mask shape.}
	\label{fig:ch4:OWFS_LBT}
	\end{center}
\end{figure}

We optimize over a frequency range $\pm 10\lambda/D$ in the frequency space. Optimizations take a lot of time because they are carried out on each point of the mask and in our basic implementation of the method, it is complicated to work with tables larger than $100\times100$. Here, we are working with 40 pixels in the pupil and twice Shannon sampling. The symmetries of the system are used to optimize on only a quarter of the focal plane. The shape of the mask thus generated and its frequency sensitivity to uniform noise are given in figure \ref{fig:ch4:OWFS_LBT}. As in the previous examples, we find the fact that the shape of the mask follows the intensity distribution in the focal plane. This optimization therefore demonstrates the proper functioning of this method for pupils with more complicated shapes than those with rotational symmetry.

This method could be applied to large segmented pupils of the ELT class, such as that of the \textit{Giant Magellan Telescope} (GMT) \cite{2014SPIE.9145E..1FJ}. It would still be necessary to make an effort in terms of improving the calculation time to be able to work with larger tables. There are two path to explore:
\begin{enumerate}
\item Reduction of parameters thanks to symmetries: in the previous case, the symmetry of revolution is not valid but for redundancy reasons, the optimization could be done on a quarter of the mask only.
\item Improvement of the algorithm: the function \textit{lsqnonlin} calculates the gradient of the score function using finite element type methods. One way to improve this process is to analytically calculate the gradient of the score function, in order to implement the result directly in the optimization algorithm. This calculation has been carried out but not yet implemented within the method.
\end{enumerate}

\subsection{Optimization with constraints}

The widely used gradient-descent algorithm also deliver constrained optimization techniques. It is then possible to optimize a score function while constraining another criterion. This can be useful for optimizing a mask characteristic while constraining high sensitivity.

Here we take an example: in the case of a full pupil, we are trying to create a mask that creates flat reference intensities over the pupil footprint while maintaining very good sensitivity. We know the expression of the reference intensities as a function of the mask: $I_{0} = |\mathbb{I}_{p}\star\widehat{m}|^{2}$. We therefore want $I_{0}$ to have the form closest to $\mathbb{I}_{p}$ possible. With a pure phase mask, the new score function is:

\begin{equation}
\mathcal{F}(\Delta) = ||\mathbb{I}_{p}-|\mathbb{I}_{p}\star\widehat{e^{i\Delta}}|^{2}||_{2}
\label{eq:ch4:Iref_score}
\end{equation}

A trivial solution that minimizes our score function is $\Delta=0$, i.e no focal plane phase mask. This is why the minimization must be constrained, so that the mask produced remains a mask with non-zero sensitivity. The constraint is written as a function $\mathcal{C}$ whose value is forced to be less than $0$ during the optimization. The constraint function $\mathcal{C}$ on the sensitivity of the mask can be written:

\begin{equation}
\mathcal{C}(\Delta) = ||4-|2\text{Im}\big(e^{i\Delta}\star e^{-i\Delta}PSF\big)|^2\star PSF||_{2} - X < 0
\label{eq:ch4:Iref_contrainte}
\end{equation}

where $X$ is a parameter that allows us to choose the level of constraint that one want to put on the mask sensitivity.\\

\begin{figure}[ht]
	\begin{center}
	\includegraphics[scale=0.7]{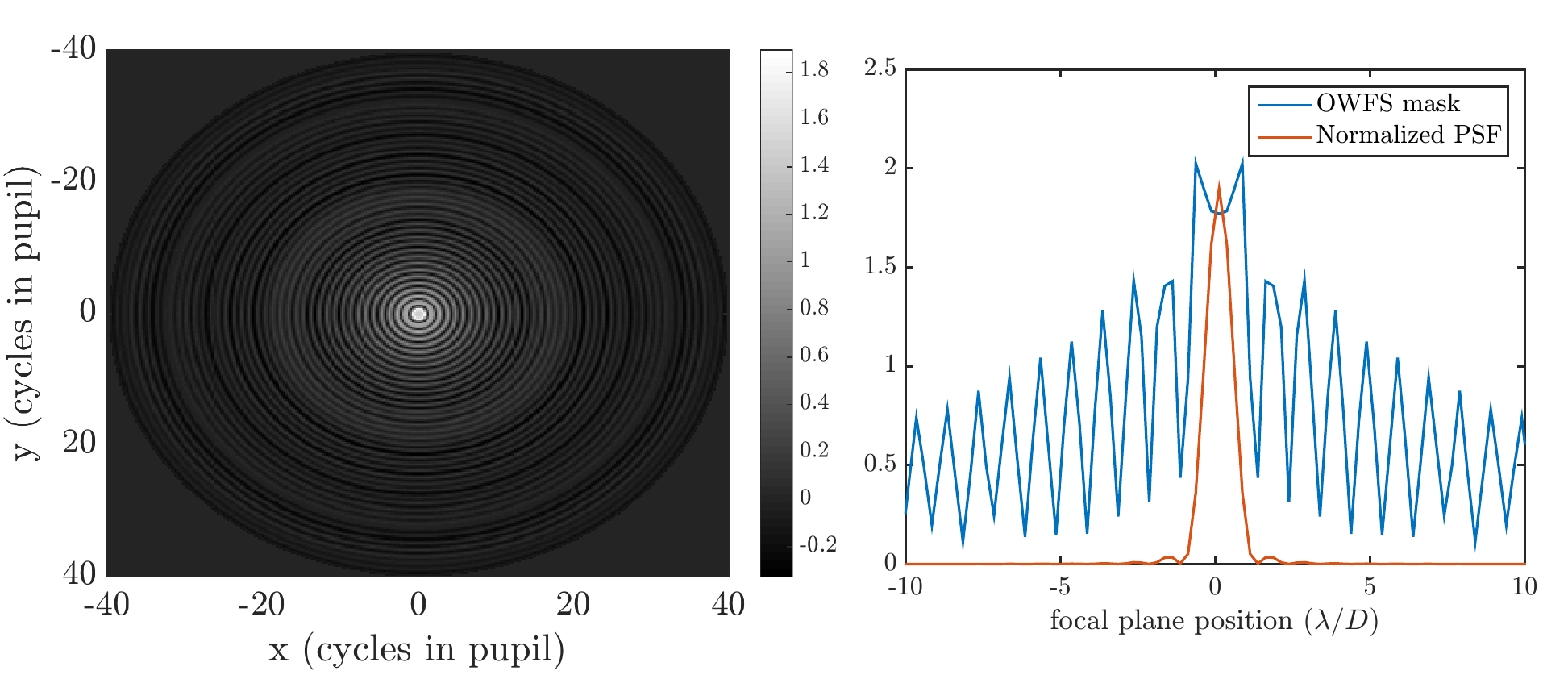}
	\caption{\textbf{Mask optimized with constraints} in order to have flat reference intensities.}
	\label{fig:ch4:AWFS}
	\end{center}
\end{figure}

This optimization is performed under constraints using the MATLAB function \textit{fmincon}. We come back to the case of the full pupil, with a symmetry of revolution. The mask produced is given in figure \ref{fig:ch4:AWFS}.

\begin{figure}[ht]
	\begin{center}
	\includegraphics[scale=0.6]{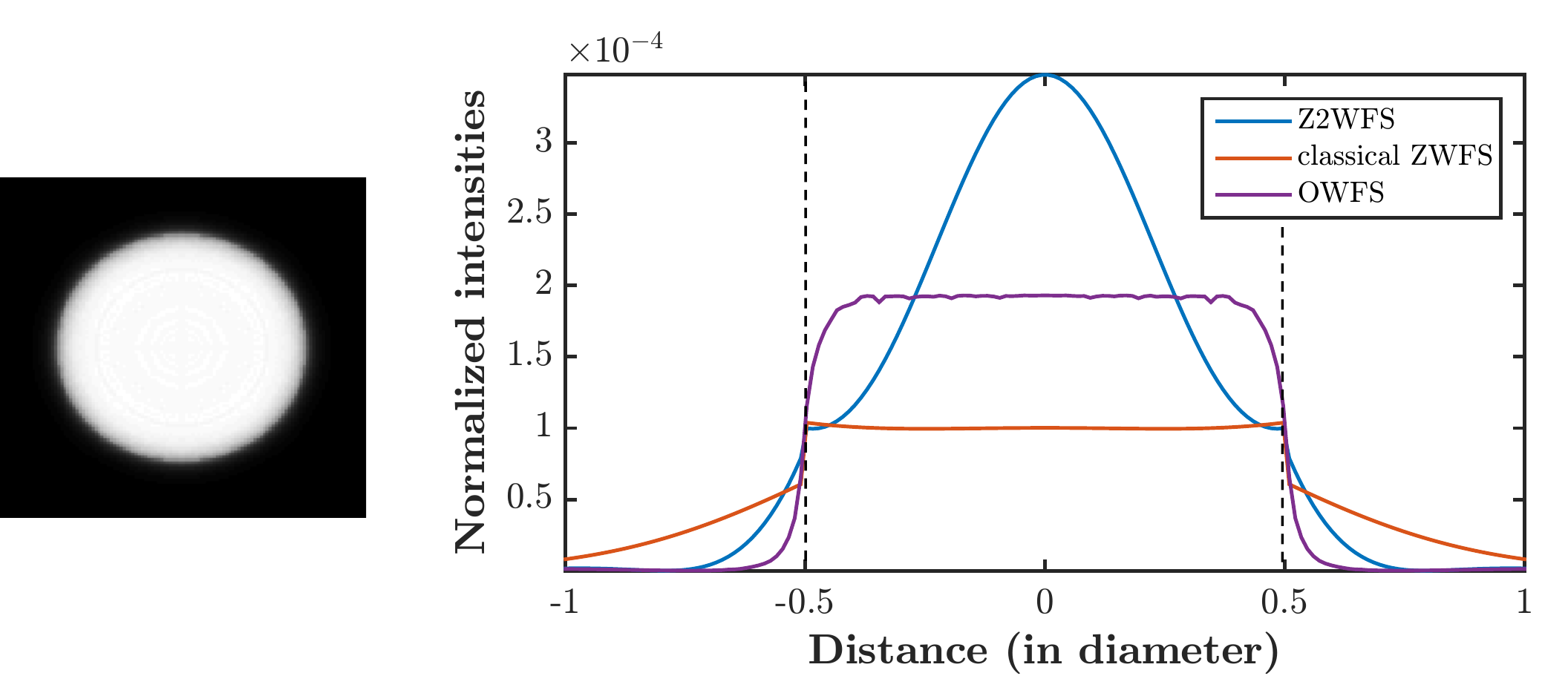}
	\caption{\textbf{Reference intensities for the sensor optimized with constrains.} pupil edges are in black dashed lines.}
	\label{fig:ch4:AWFS_Iref_cut}
	\end{center}
\end{figure}

The mask presents a depletion in the center, with quite substantial rebounds. The reference intensities are presented in figure \ref{fig:ch4:AWFS_Iref_cut}. The optimization has indeed converged towards a mask whose reference intensities are very uniform.

\begin{figure}[ht]
	\begin{center}
	\includegraphics[scale=0.6]{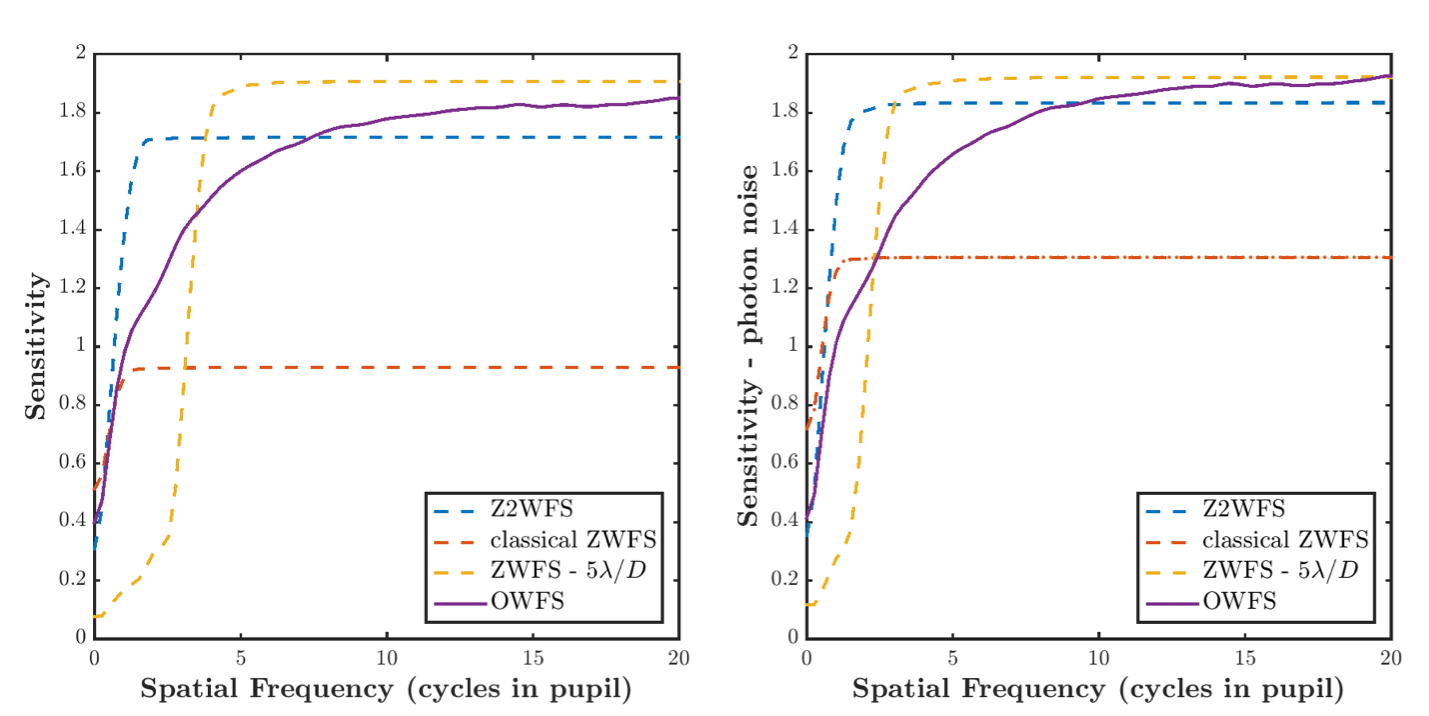}
	\caption{\textbf{Sensitivities for the mask optimized with constrains}.}
	\label{fig:ch4:AWFS_sensi}
	\end{center}
\end{figure}

It is now necessary to ensure that the sensitivity has been little affected. This is confirmed with the sensitivity curves shown in figure \ref{fig:ch4:AWFS_sensi}. This example gives a good overview of the possibilities offered by this kind of optimization, which allows a large control of the characteristics of the mask.

\section{Conclusions and perspectives}

In this paper, a new approach to create FFWFS is proposed. It consists in optimizing the masks via non-linear optimization techniques applied on a criteria based on a convolutionnal model. New masks that closely match the fundamental sensitivity limits have been created. These masks can been described as "larger and apodized ZWFS", whose shapes are adapted to the entrance pupil. This study was only focusing on sensitivity in the linear regime. These new FFWFS have actually the same drawbacks as the ZWFS class: high chromaticity and reduced dynamic range. Therefore, they could be well suited as second stage AO WFS for High Contrast Imaging applications. A practical implementation of these masks could be done through liquid-crystal masks which are achromatic in phase-shift as done in \cite{Doelman2018} for the ZWFS. Besides, building achromatic mask in phase is already an explored area in the field of cornography \cite{Mawet:09}. It will remains the issue of the radial chromaticity, harder to solve in a practical way.

The tool developed therefore allows us to have great control over the sensitivity of the masks. But the optimization was done on the RON sensitivity only. We could improve this technique to optimize on the photon noise sensitivity. It could be done thanks to the optimization with constrains, which could allow to set low reference intensities. On top of that, we would logically like to be able to optimize the masks also in terms of their dynamics. It seems very promising to try to build WFS with high dynamics while having a high sensitivity, still by using optimization with constraints.

\acknowledgments 
 
This work benefited from the support of the WOLF project ANR-18-CE31-0018 of the French National Research Agency (ANR). It has also been prepared as part of the activities of OPTICON H2020 (2017-2020) Work Package 1 (Calibration and test tools for AO assisted E-ELT instruments). OPTICON is supported by the Horizon 2020 Framework Programme of  the  European  Commission’s  (Grant  number  730890). Authors are acknowledging the support by the Action Spécifique Haute Résolution Angulaire (ASHRA) of CNRS/INSU co-funded by CNES. Vincent Chambouleyron PhD is co-funded by "Région Sud" and ONERA, in collaboration with First Light Imaging. Finally, part of this work is supported by the LabEx FOCUS ANR-11-LABX-0013.

\bibliography{report} 
\bibliographystyle{spiebib} 

\end{document}